\documentclass{iopjournalarXiv}
\usepackage{amsmath, natbib, subcaption}
\setcitestyle{numbers,square,comma,sort&compress}

\begin{document}

\articletype{article} 

\title{Efficient calculation of real-space lattice propagators in the presence of a Fermi sea}

\author{Oliver Tong$^{1,2,*}$\orcid{0009-0008-1845-1631} and  Mona Berciu$^{1,2}$\orcid{0000-0002-6736-1893}} 

\affil{$^1$Department of Physics and Astronomy, University of British Columbia, Vancouver, Canada}

\affil{$^2$Quantum Matter Institute, University of British Columbia, Vancouver, Canada}

\affil{$^*$Author to whom any correspondence should be addressed.}

\email{ot83@cornell.edu} %% TODO: replace with Oliver's email address


\begin{abstract}
We present an efficient method for computing the real-space propagators (lattice Green's functions) of any tight-binding Hamiltonian in the presence of a Fermi sea with carrier concentration $x$. The method is valid for any lattice, dispersion, and dimension, provided the corresponding $x=0$ propagators are known. We show that suitable combinations of the finite-$x$ particle-addition propagators have a real or imaginary part trivially related to their $x=0$ counterpart, while the remaining part follows from a Kramers-Kronig relation that can be evaluated for all energies at once using the fast Fourier transforms. The computational cost is therefore independent of the dimensionality, unlike that of direct Brillouin-zone integration. The particle-removal propagators follow from general identities. We validate the method against direct integration for hypercubic lattices in one, two, and three dimensions, and use the 2D square lattice to illustrate how the shape of the Fermi surface is imprinted on the spatial structure of the propagators. In particular, at energies far outside the band, the propagator maps converge to the Fraunhofer diffraction pattern whose aperture is the unoccupied part of the Brillouin zone.
\end{abstract}

\section{Introduction}

Consider any non-interacting, tight-binding Hamiltonian ${\cal H}=\sum_\mathbf{k} \epsilon_\mathbf{k} c^\dagger_\mathbf{k} c_\mathbf{k}$ defined on a lattice, with $\epsilon_\mathbf{k}$ the corresponding dispersion while $c^\dagger_\mathbf{k}=\sum_\mathbf{n}\frac{e^{i \mathbf{k}\cdot \mathbf{R}_\mathbf{n}}}{\sqrt{N}}c^\dagger_\mathbf{n}$ creates an electron with momentum $\mathbf{k}$ and $c^\dagger_\mathbf{n}$ creates an electron at lattice site $\mathbf{R}_\mathbf{n}$. For simplicity, we assume that the spin is a trivial degree of freedom and we ignore it. Let $|\mathrm{FS\rangle}= \prod_{\epsilon_\mathbf{k}\le E_F}c^\dagger_\mathbf{k}|0\rangle$ be the ground state for a given concentration $x$ of electrons (per spin component), consisting of a Fermi sea of  filled states up to the appropriate Fermi energy $E_F$. Finally, let $E_\mathrm{FS} = \sum_{\epsilon_\mathbf{k}\le E_F} \epsilon_\mathbf{k}$ be the corresponding ground-state energy.

In momentum space, the bare propagator is defined as \cite{Mahan}:
\begin{equation}
G(\mathbf{k}, z; x) =  G^A(\mathbf{k}, z; x) +  G^R(\mathbf{k}, z^*; x)=\langle \mathrm{FS}| c_\mathbf{k} \frac{1}{z- {\cal H}+ E_\mathrm{FS}}     c^\dagger_\mathbf{k}|  \mathrm{FS}\rangle + \langle \mathrm{FS}| c^\dagger_\mathbf{k} \frac{1}{z^* + {\cal H}- E_\mathrm{FS}}     c_\mathbf{k}|  \mathrm{FS}\rangle,
\end{equation}
where the first/second terms are the particle addition ($A$)/removal ($R$) parts; these superscripts should not be confused with the advanced/retarded propagators. Also, $z=\omega+i\eta, z^* = \omega-i \eta$ is the energy of interest together with a small artificial broadening $\eta \to 0^+$. For any such model, these propagators are trivial to calculate, {\em e.g.} the particle addition part is 
\begin{equation}
    G^A(\mathbf{k}, z; x)=  \frac{\Theta(\epsilon_\mathbf{k}-E_F)}{z- \epsilon_\mathbf{k}},
\end{equation}
where the Heaviside function shows that a fermion can only be injected above the Fermi sea. So as long as that condition is satisfied, the addition propagator $1/(z-\epsilon_\mathbf{k})$ is identical to that of a particle injected into the empty system ($x=0$), reflecting the lack of interactions in the model.

In real-space, and taking advantage of invariance to translations, the bare propagators are:
\begin{equation}
G_\mathbf{n}(z; x) = G^A_\mathbf{n}(z; x) +G^R_\mathbf{n}(z^*; x) =  \langle \mathrm{FS}| c_\mathbf{n} \frac{1}{z- {\cal H}+ E_\mathrm{FS}}     c^\dagger_\mathbf{0}|  \mathrm{FS}\rangle + \langle \mathrm{FS}| c^\dagger_\mathbf{n} \frac{1}{z^* + {\cal H}- E_\mathrm{FS}}     c_\mathbf{0}|  \mathrm{FS}\rangle,
\end{equation}
For reasons that will soon become apparent, we will focus from now on the particle addition part. This can be expressed as 

\begin{equation}
      G^A_{\mathbf{n}}(z;x)= \frac{1}{N}\sum_{\mathbf{k}} e^{i\mathbf{k}\cdot \mathbf{R}_\mathbf{n}}\frac{\Theta(\epsilon_{\mathbf{k}} - E_F)}{z-\epsilon_{\mathbf{k}}}
      \label{4}
\end{equation}
 Unlike for the momentum-space propagator, this depends non-trivially on $E_F$ (and thus, on $x$), since it controls which momenta contribute to the total interference pattern that defines this amplitude of probability to find at site $\mathbf{n}$ a particle originally injected at the origin.

 The ability to efficiently calculate these real-space propagators is essential for dealing with problems where translational invariance is broken, {\em e.g.} by the presence of impurities, disorder and/or of surfaces \cite{economou2006greens, Einstein, balatsky2006impurity, lopezsancho1985highly}, as well as when describing systems where quasi-particles are dressed by clouds more conveniently described in real space \cite{bonca1999holon, berciu2006green, goodvin2006green}.  

 For a 1D chain with nearest-neighbour hopping, these real-space propagators can be calculated analytically~\cite{berciu2022polarons}, but any other case requires numerical computation. From Eq. (\ref{4}), it is clear that the real-space addition propagator can be calculated as a $D$-dimensional integral over the part of the Brillouin zone that lies outside the Fermi sea. However, this is not an efficient approach, especially in higher dimensions. In fact, even for $x=0$, there has been a lot of effort to find ways to calculate the real-space propagators which avoid direct integration. These include methods based on recurrence relations \cite{morita1971useful}, continued fractions \cite{berciu2009computing, berciu2010efficient}, and the ‘Chebyshev analytic continuation’ method \cite{loh2017general}, which derives the Chebyshev coefficients and power series of these propagators based on the number of paths connecting sites $\mathbf{n}$ and $\mathbf{0}$. More recently, improvements were made to the large-distance approximation and power series expansion of these $x=0$ propagators, by combining these improvements together a general recurrence relation, a composite algorithm is developed which can efficiently compute propagators at any energy and distance for lattices with root-free dispersions \cite{zhugayevych2025efficient}.

In this article, we present a very efficient way to calculate the finite-$x$ real-space lattice propagators for any lattice and any tight-binding model in any dimension, if the $x=0$ real-space propagators are known. We exemplify it for a 2D square lattice  with nearest-neighbour hopping simply because it is easier to show the real-space maps, but the numerical effort is the same for any tight-binding model on any lattice in any dimension (once the corresponding $x=0$ propagators are available).

\section{Method}
In the thermodynamic limit, Eq. (\ref{4}) becomes:
\begin{equation}
    G^A_{\mathbf{n}}(z;x)  = \frac{1}{(2\pi)^d}\int_{BZ} d^d\mathbf{k} \frac{e^{i\mathbf{k}\cdot \mathbf{R}_\mathbf{n}}}{z-\epsilon_{\mathbf{k}}}\Theta(\epsilon_{\mathbf{k}} - E_F),\label{addsum}
\end{equation}
and the removal part is:
\begin{equation}
     G^R_{\mathbf{n}}(z;x) = \frac{1}{(2\pi)^d}\int_{BZ} d^d\mathbf{k} \frac{
    e^{-i\mathbf{k}\cdot \mathbf{R}_\mathbf{n}}}{z-\epsilon_{\mathbf{k}}}\Theta(E_F - \epsilon_{\mathbf{k}}). \label{removeintegral}
\end{equation}
As a result:
\begin{equation} \label{GeneralARrelation}
    G^R_{-\mathbf{n}}(z;x) = G^A_{\mathbf{n}}(z;0) - G^A_{\mathbf{n}}(z;x).
\end{equation}
Furthermore, if the dispersion relation has a symmetry such that there is some $\mathbf{Q}$ for which $\epsilon_{\mathbf{Q}-\mathbf{k}} = -\epsilon_{\mathbf{k}}$ for all  $\mathbf{k}$, then 
\begin{equation}\label{bipartiteARrelation}
    G^A_{\mathbf{n}}(z;1-x) = -e^{i\mathbf{Q}\cdot\mathbf{R}_\mathbf{n}} G^R_{\mathbf{n}}(-z;x).
\end{equation}
For our 2D example below, this holds true for $\mathbf{Q}= (\pi, \pi)$. We note that Eq. (\ref{bipartiteARrelation}) is a generalization of the relation between the fermion addition and removal propagators reported in 1D in \cite{berciu2022polarons}. Due to these relations between $G^R$ and $G^A$, from now on we focus on the latter only. 

Using the identity $\lim_{\eta\to 0^+} \frac{1}{\omega+i\eta - \epsilon_\mathbf{k}} = \mathcal{P}\frac{1}{\omega-\epsilon_\mathbf{k}}-i\pi \delta(\omega-\epsilon_\mathbf{k})$,  we can formally take the limit $\eta \to 0$ in Eq.  (\ref{addsum}),  to find:
\begin{multline}\label{addintexpand}
    G^A_{\mathbf{n}}(\omega; x) =  \frac{1}{(2\pi)^d}\int_{BZ} d^d\mathbf{k}
    \Big[\mathcal{P}\frac{1}{\omega-\epsilon_\mathbf{k}}\cos(\mathbf{k}\cdot \mathbf{R}_\mathbf{n})+\pi \delta(\omega-\epsilon_\mathbf{k})\sin(\mathbf{k}\cdot \mathbf{R}_\mathbf{n})\Big]\Theta(\epsilon_{\mathbf{k}}-E_F)  \\ + \frac{i}{(2\pi)^d}\int_{BZ} d^d\mathbf{k}
    \Big[\mathcal{P}\frac{1}{\omega-\epsilon_\mathbf{k}}\sin(\mathbf{k}\cdot \mathbf{R}_\mathbf{n})-\pi \delta(\omega-\epsilon_\mathbf{k})\cos(\mathbf{k}\cdot \mathbf{R}_\mathbf{n})\Big]\Theta(\epsilon_{\mathbf{k}}-E_F)
\end{multline}

Any integral of the form $I(\omega)=\int_{BZ} d^d\mathbf{k} f(\mathbf{k})\delta(\omega-\epsilon_{\mathbf{k}})\Theta(\epsilon_{\mathbf{k}}-E_F)$ with $f$ being a smooth function, vanishes if $\omega<E_F$, while for any $\omega> E_F$ the result is the same as when $x=0$. The two such terms from Eq.~(\ref{addintexpand}) can be isolated by using:
\begin{align}\nonumber
    \frac{1}{2}\mathrm{Re}\Big[G^A_{\mathbf{n}}(\omega;x) - G^A_{-\mathbf{n}}(\omega; x)\Big]=&\frac{\pi}{(2\pi)^d}\int_{BZ} d^d\mathbf{k} \delta(\omega-\epsilon_\mathbf{k})\sin(\mathbf{k}\cdot \mathbf{R}_\mathbf{n})\Theta(\omega-E_F)\\
    =& \frac{\Theta(\omega-E_F)}{2}\mathrm{Re}\Big[G^A_{\mathbf{n}}(\omega;0) - G^A_{-\mathbf{n}}(\omega; 0)\Big] \label{intreal}
\end{align}
and
\begin{align}\nonumber
    \frac{1}{2}\mathrm{Im}\Big[G^A_{\mathbf{n}}(\omega; x) + G^A_{-\mathbf{n}}(\omega; x)\Big]=& -\frac{\pi}{(2\pi)^d}\int_{BZ} d^d\mathbf{k} \delta(\omega-\epsilon_\mathbf{k})\cos(\mathbf{k}\cdot \mathbf{R}_\mathbf{n})\Theta(\omega-E_F) \\
    = & \frac{\Theta(\omega-E_F)}{2}\mathrm{Im}\Big[G^A_{\mathbf{n}}(\omega; 0) + G^A_{-\mathbf{n}}(\omega; 0)\Big]. \label{intimag}
\end{align}
Therefore, at any concentration $x$, Eqs. (\ref{intreal}) and (\ref{intimag}) give the real (imaginary) part of the difference (sum) of the two propagators in terms of their corresponding $x=0$ values. The finite-$x$ imaginary (real) part of the same combinations can then be obtained by applying Kramers-Kronig relations onto Eqs. (\ref{intreal}) and (\ref{intimag}). The propagators  $G^A_{\mathbf{n}}(\omega; x)$ and $G^A_{-\mathbf{n}}(\omega; x)$ are then obtained straightforwardly.

The Kramers-Kronig relation can be applied over all energies at once by expressing it in Fourier space using the Hilbert transform:

\begin{equation}
    \mathcal{F}\Big[\mathrm{Im}\big(G^A_{\mathbf{n}}(\omega; x) - G^A_{-\mathbf{n}}(\omega; x)\big)\Big](f)=-i\, \mathrm{sgn}(f)\, \mathcal{F}\Big[\mathrm{Re}\big(G^A_{\mathbf{n}}(\omega; x) - G^A_{-\mathbf{n}}(\omega; x)\big)\Big](f)
\end{equation}
\begin{equation}
    \mathcal{F}\Big[\mathrm{Re}\big(G^A_{\mathbf{n}}(\omega; x) + G^A_{-\mathbf{n}}(\omega;x)\big)\Big](f)=i\, \mathrm{sgn}(f)\, \mathcal{F}\Big[\mathrm{Im}\big(G^A_{\mathbf{n}}(\omega; x) + G^A_{-\mathbf{n}}(\omega;x)\big)\Big](f),
    \label{13}
\end{equation}
where $\mathcal{F}[g](f) = \int d\omega\, e^{-2\pi i f \omega}\, g(\omega)$ denotes the Fourier transform from energy $\omega$ to its conjugate variable $f$. 

If the lattice has inversion symmetry, then  $G^A_\mathbf{n}(\omega; x) = G^A_\mathbf{-n}(\omega; x)$ and Eqs. (\ref{intreal})- (\ref{13}) further simplify to  $\mathrm{Im} G^A_\mathbf{n}(\omega; x)=\Theta(\omega-E_F) \mathrm{Im} G^A_\mathbf{n}(\omega; 0)$, and $\mathrm{Re} G^A_\mathbf{n}(\omega; x)$ can then be computed using Kramers-Kronig, see Eq. (\ref{13}),  with no need to also separately compute $G^A_\mathbf{-n}(\omega; x)$. This allows us to more clearly see how the finite-$x$ affects the real-space propagators. Their imaginary part (linked to local  densities of states) becomes zero for $\omega< E_F$, reflecting the Fermi blockade from the other electrons, and is unchanged for $\omega> E_F$, reflecting the lack of interactions. However, the real part changes at all $\omega$ as a function of $x$. For $\omega$ within the band, this change is generally more significant for $\omega \sim E_F$ and for smaller values of $\mathbf{n}$.

Finally, we note that this method computes $G_\mathbf{n}^A( \omega; x)$ strictly in the limit $\eta =0$. To obtain $G_\mathbf{n}^A(\omega + i\eta; x)$ for a finite $\eta$, the Lorentzian broadening is applied by convolving the $\eta=0$ result  with a Lorentzian Kernel of width $\eta$.

\begin{figure}[t]
    \centering
\includegraphics[width=0.32\linewidth]{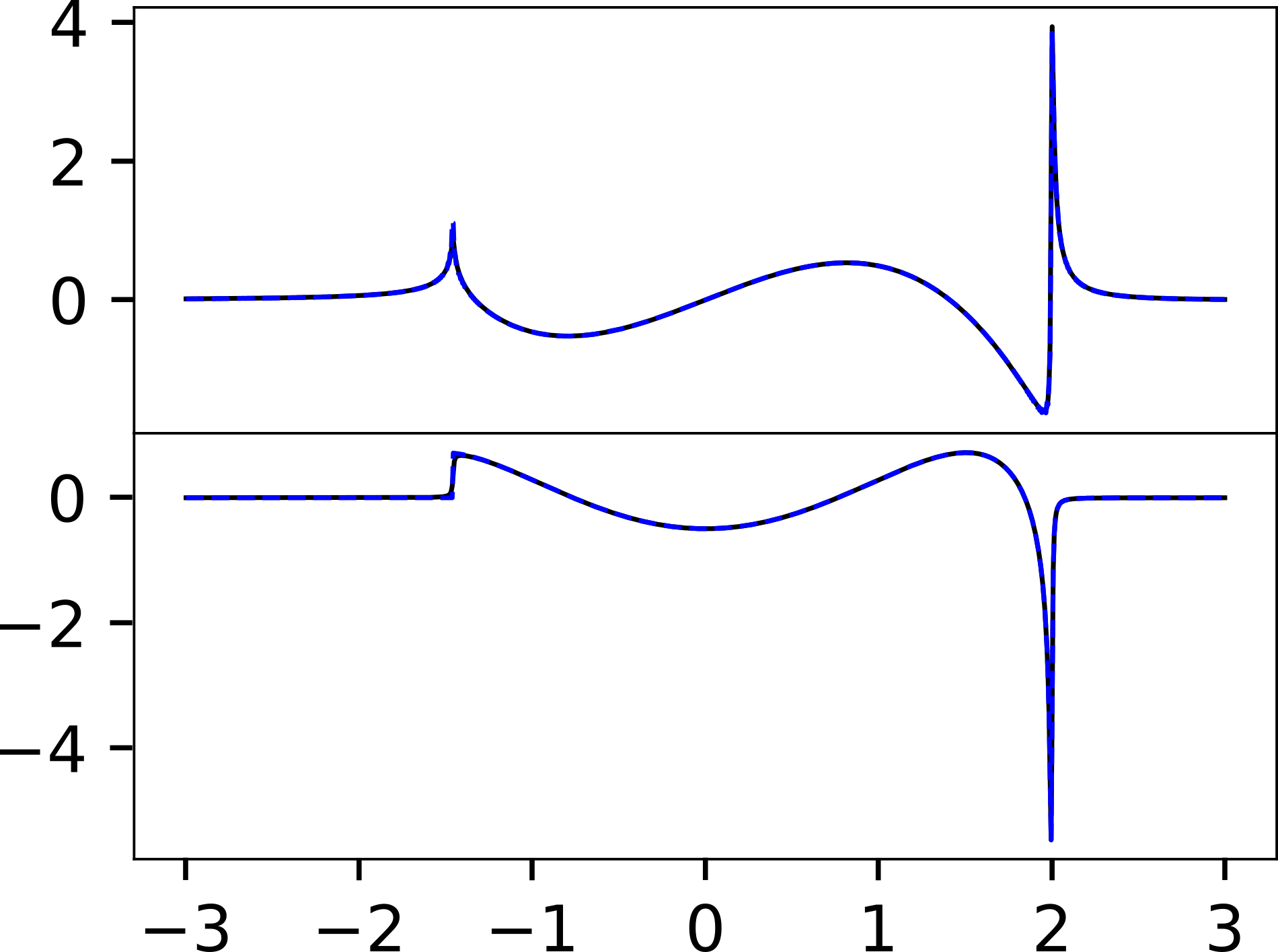}
\includegraphics[width=0.32\linewidth]{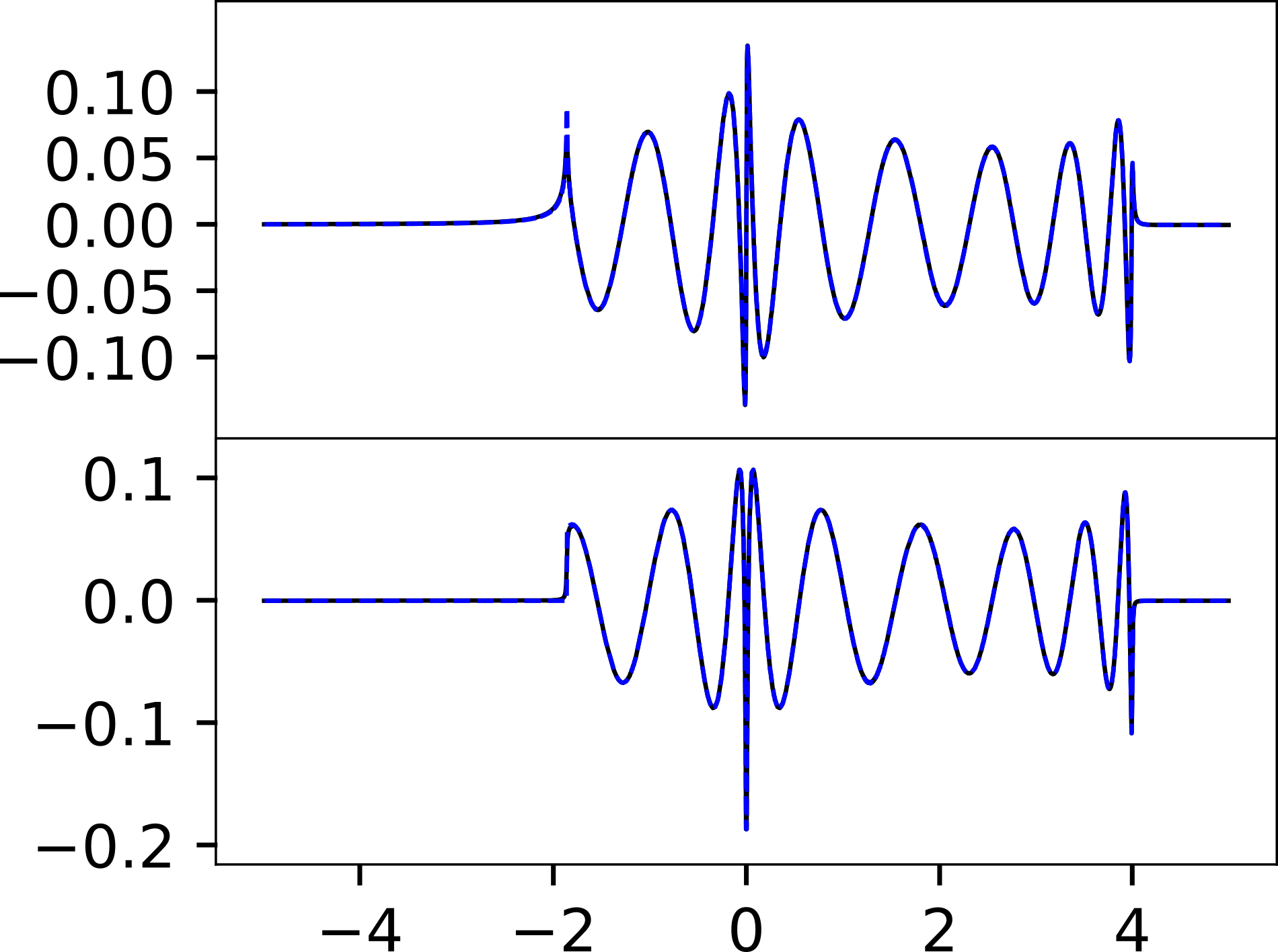}
\includegraphics[width=0.32\linewidth]{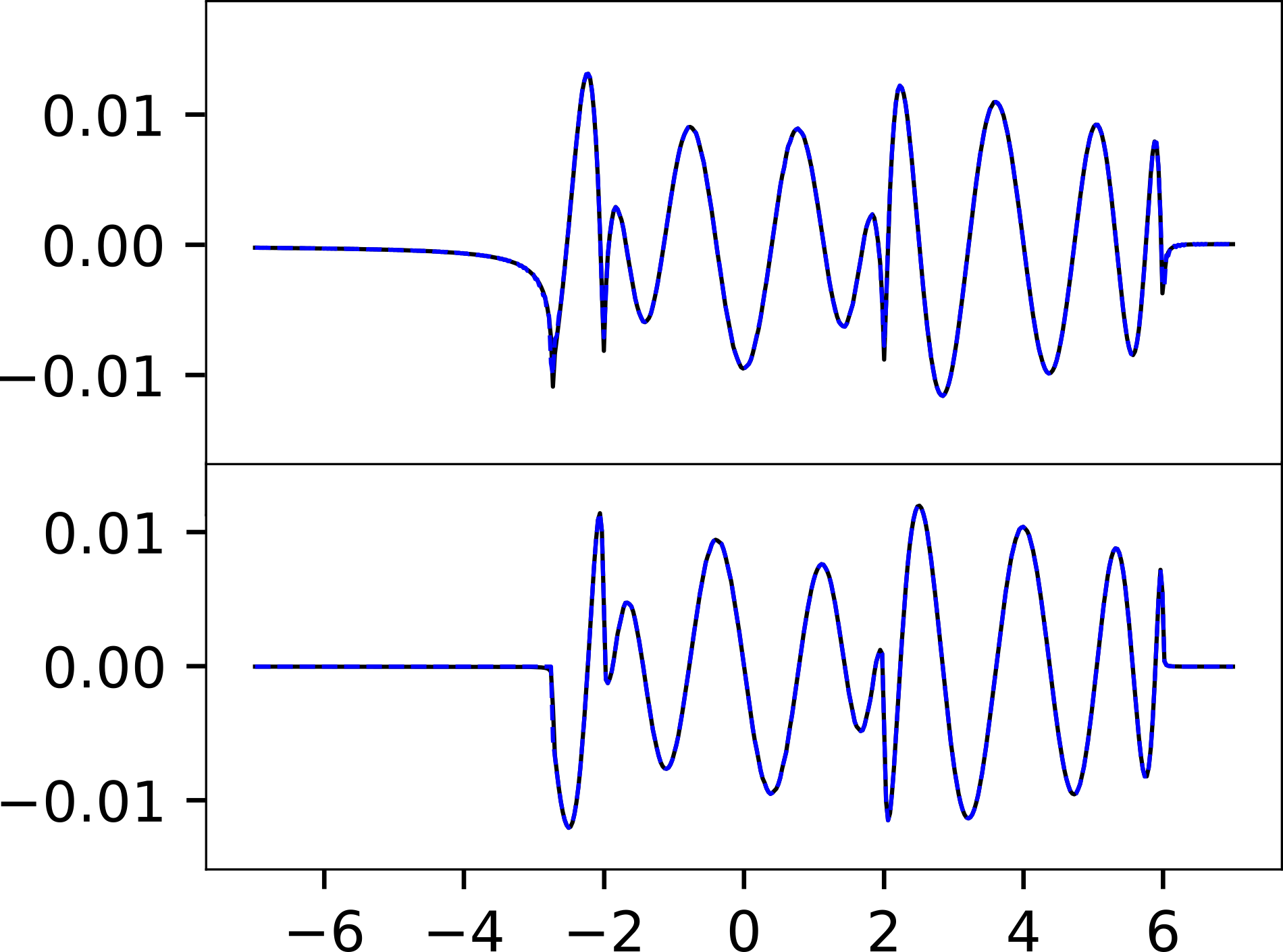}
\caption{Real (top) and imaginary (bottom) components of $G_\mathbf{n}(z; x)$ at $\eta = 0.005$ computed by direct integration (black solid curve) and via our method (blue dashed curve) in (left) 1D at $n=4, x=0.24$; (middle) 2D at $\mathbf{n} = (12,8), x=0.2$; and (right) 3D at $\mathbf{n} = (3, 8, 4), x=0.14$. In all cases $t=1$.}
    \label{fig:gvsE}
\end{figure}

\section{Results}
The method presented above is general and valid for any model with discrete translational symmetry.  It leads to a very significant numerical speedup because once $G^A_\mathbf{n}(\omega; 0)$ is known, any $G^A_\mathbf{n}(\omega; x)$ can be calculated with  Kramers-Kronig at any $x$, and a single energy integral over a known, finite energy range needs to be performed independent of $D$. 

Figure  \ref{fig:gvsE} shows a few illustrative comparisons between real-space propagators calculated with this method versus direct momentum integration, for hypercubic lattices in $D=1,2,3$ and for a nearest-neighbor hopping where we set $t=1$.

From now on we focus on the 2D square lattice with nearest-neighbour hopping and use it to illustrate some of the  effects of the Fermi sea, and in particular of the shape of the Fermi surface,  on the real-space lattice propagators. 
%We note that the analytic relations shown below are general to any lattice with inversion symmetry and a $\mathbf{Q}$ where $\epsilon_{\mathbf{Q}-\mathbf{k}} = -\epsilon_\mathbf{k}$. For the 2D square lattice, $\mathbf{Q} = (\pi, \pi)$.

We begin by considering the behavior of the propagators as a function of $x$ at energies below the free electron bandwidth. For our 2D example, we illustrate this at $\omega=-10t$ in the next few plots; the behaviour is qualitatively similar at other values.

\begin{figure}[t]
\includegraphics[width=0.24\linewidth]{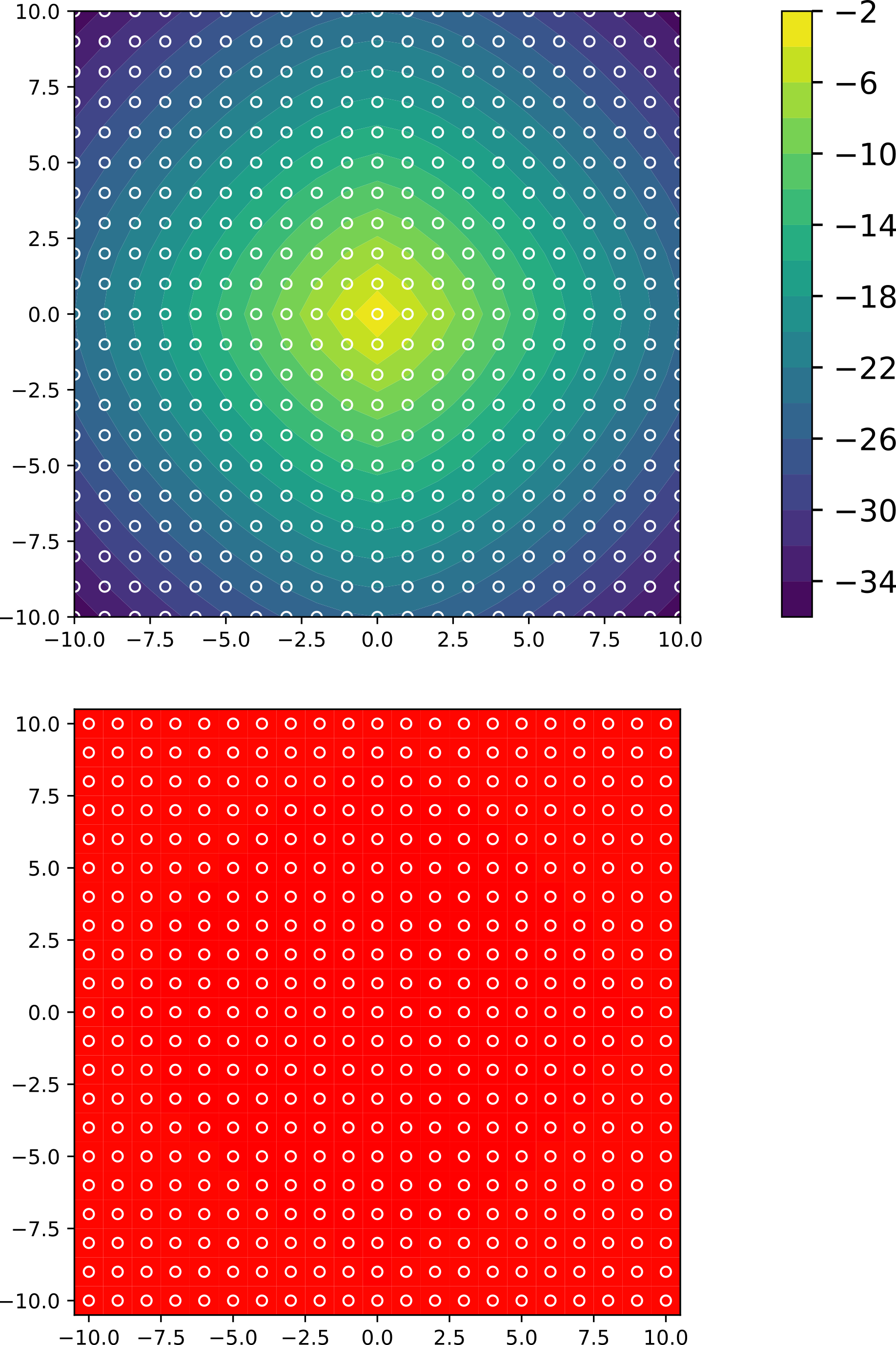}
\includegraphics[width=0.24\linewidth]{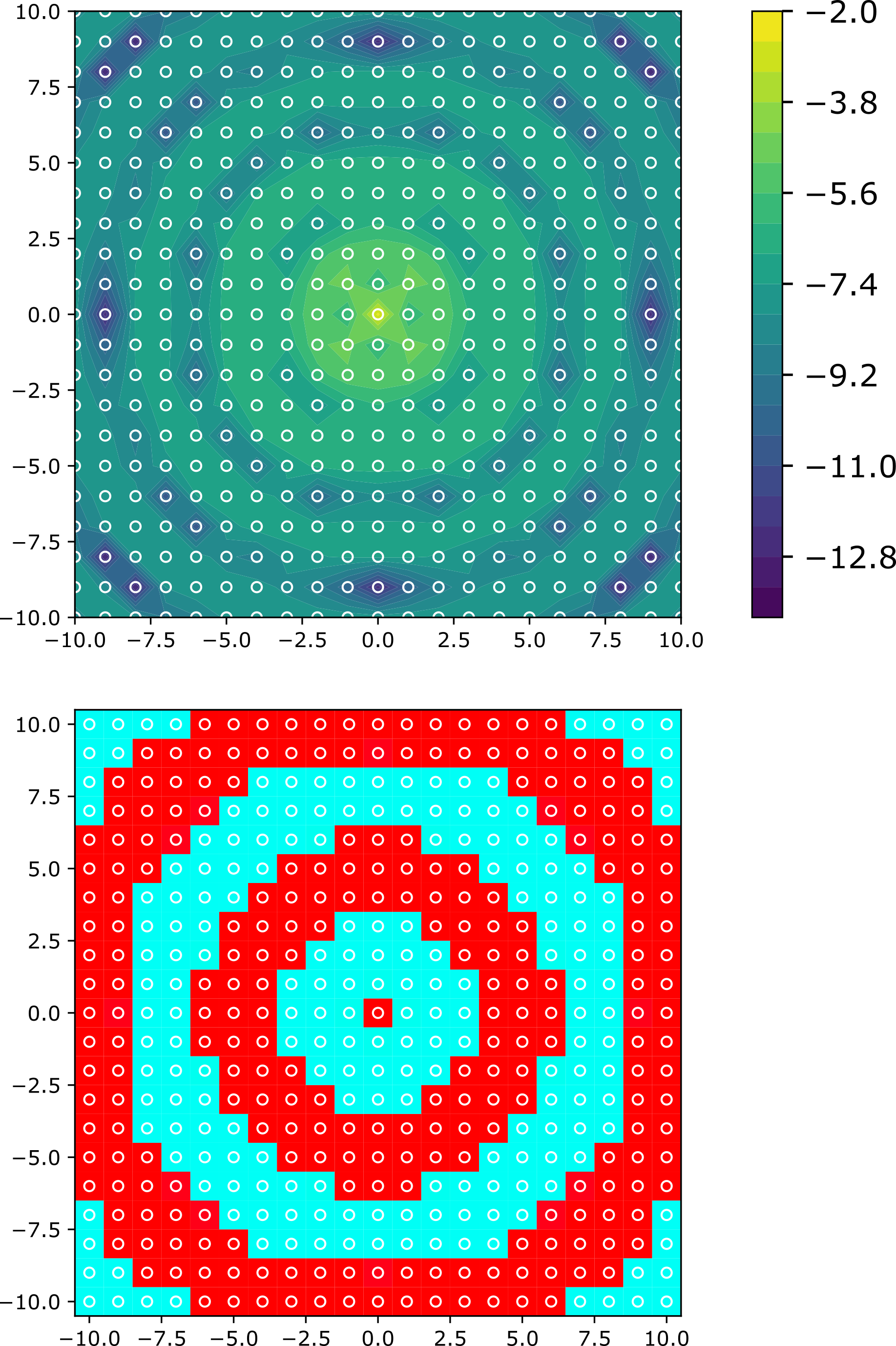}
\includegraphics[width=0.24\linewidth]{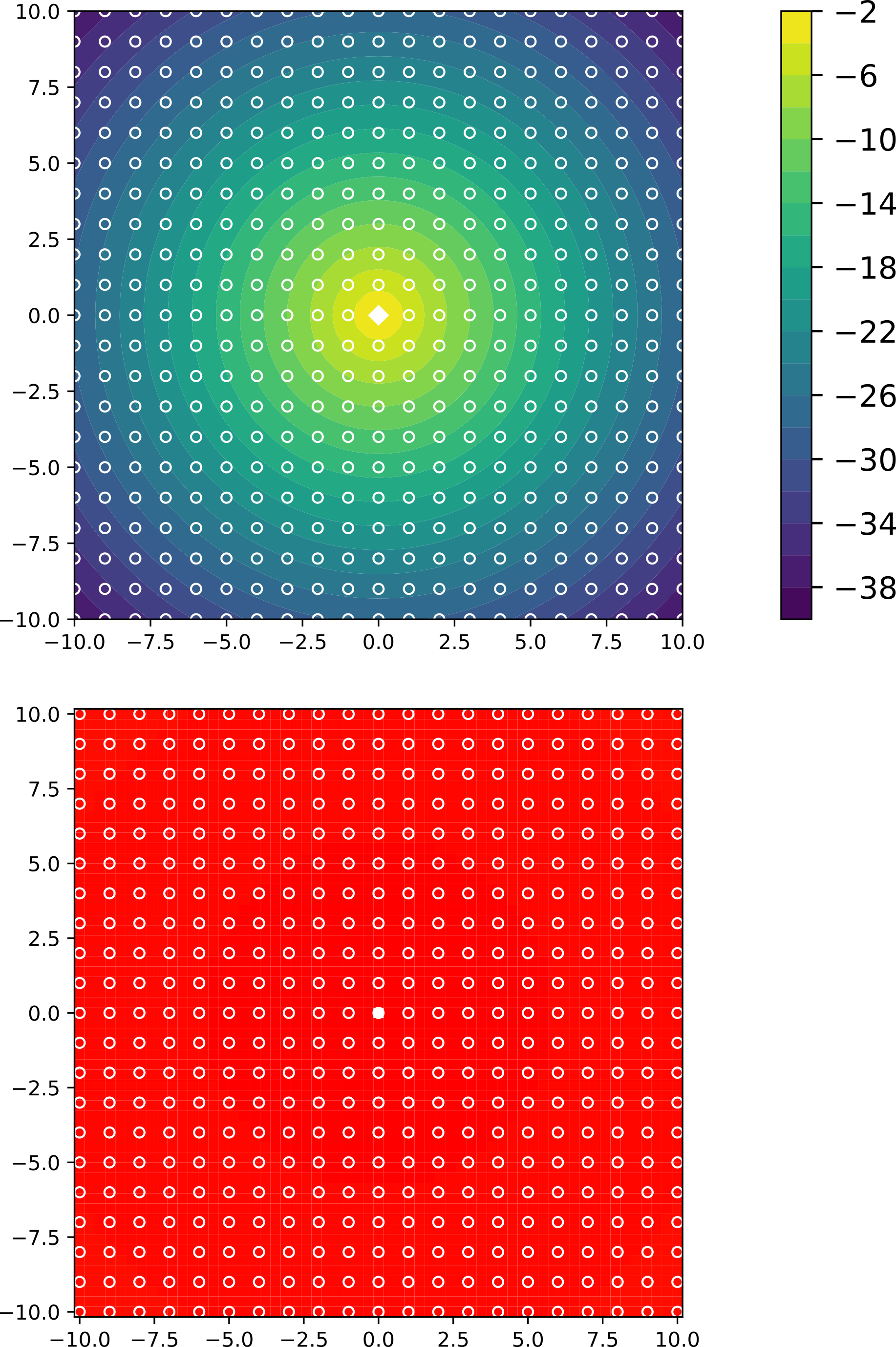}
\includegraphics[width=0.24\linewidth]{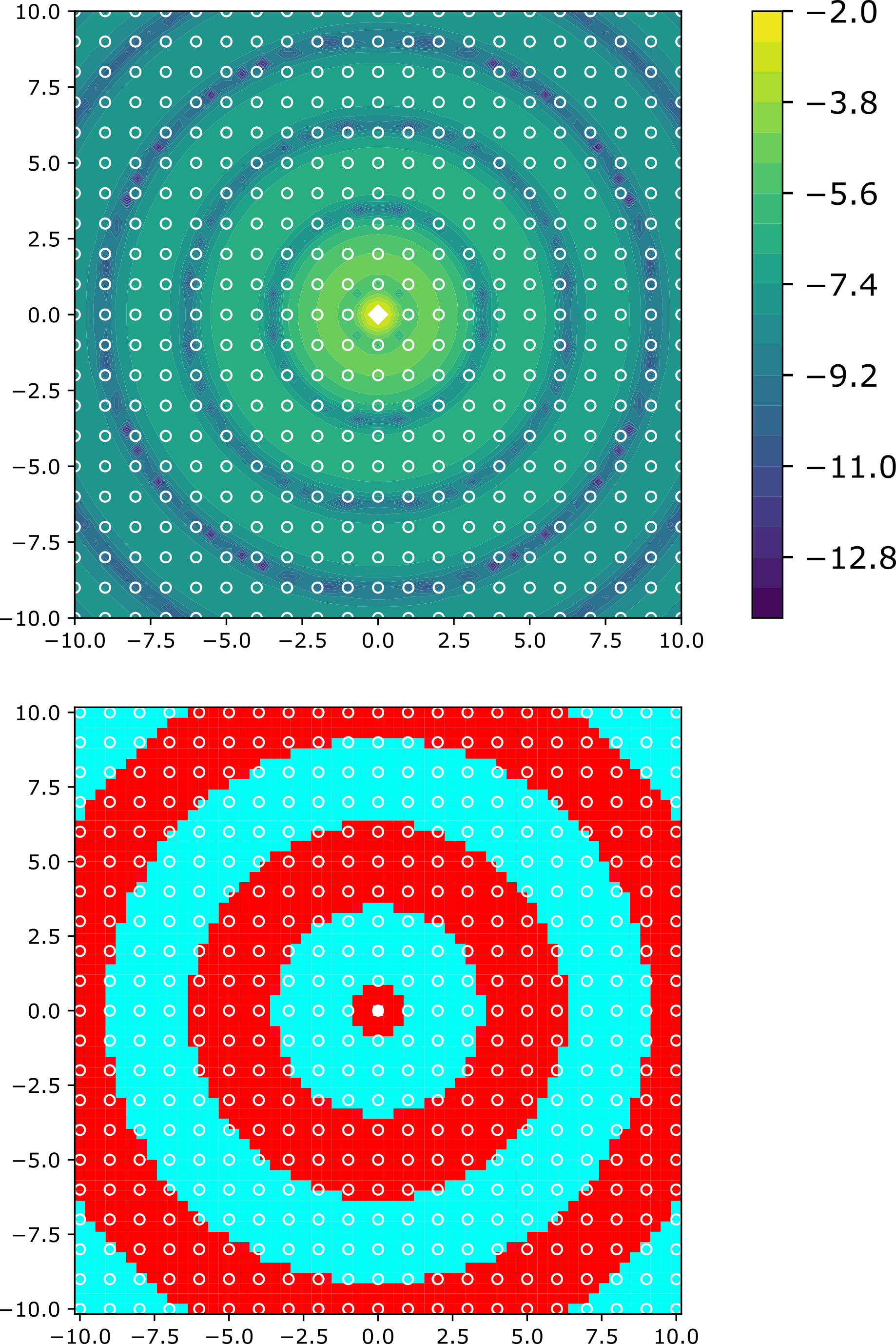}
\caption{Upper row: Maps of $\log(|G^A_\mathbf{n}(-10+0.02i; x)|)$,  and Lower row: Maps of its overall sign, with red/teal showing $G^A<0$ and $G^A>0$, respectively. The lattice sites $\mathbf{n}$ are indicated by small white circles, and only the values at these points are meaningful; the heat map is simply to help visualize the changes with $\mathbf{n}$. The left columns show the lattice results for $x=0$ and $x=0.1$, respectively. The right columns show the results of the continuum approximation for the same values. In all cases, the hopping $t=1$. }
    \label{fig:x=0}
\end{figure}

\begin{figure}[b]
\includegraphics[width=0.24\linewidth]{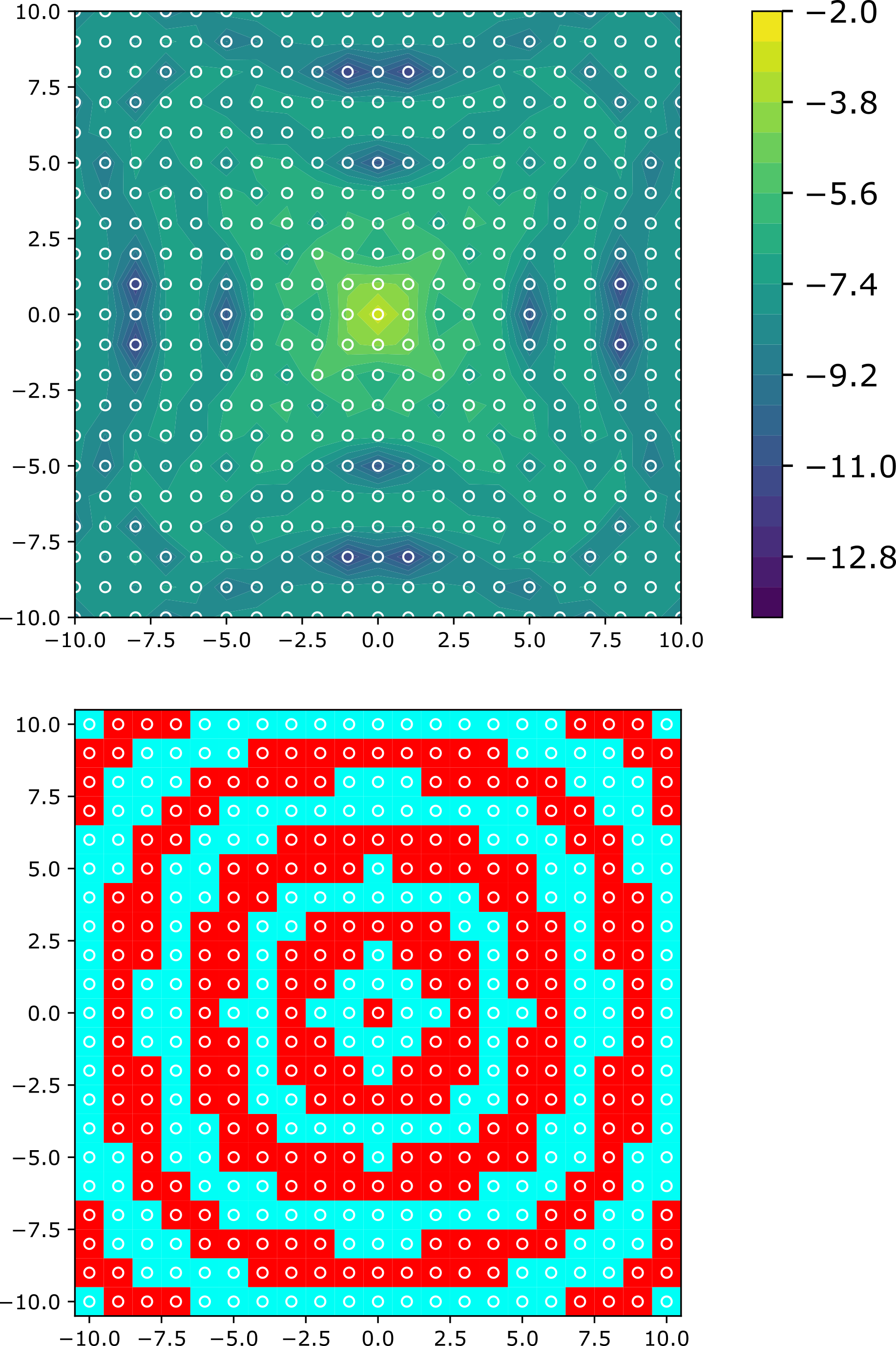}
\includegraphics[width=0.24\linewidth]{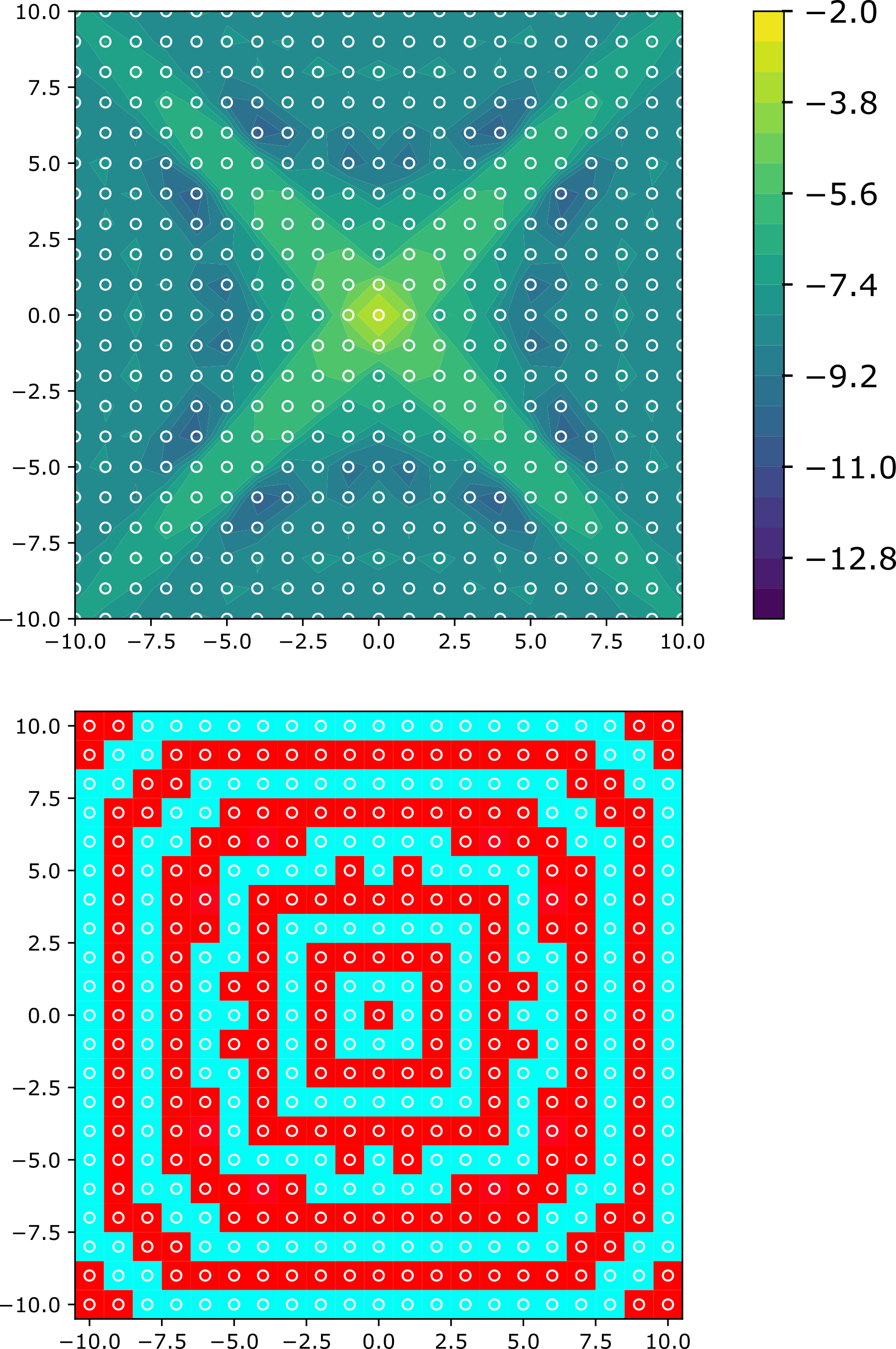}
\includegraphics[width=0.24\linewidth]{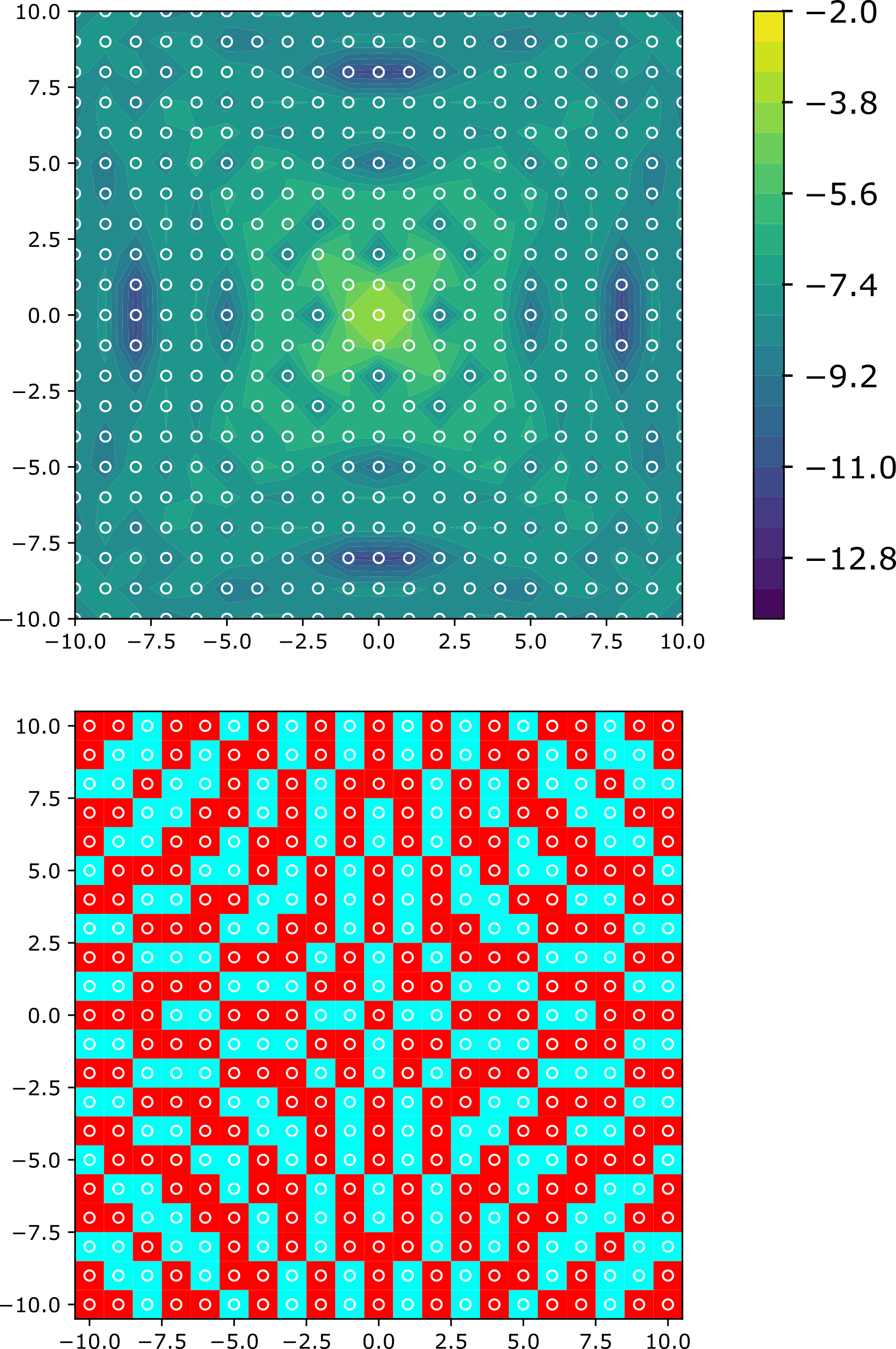}
\includegraphics[width=0.24\linewidth]{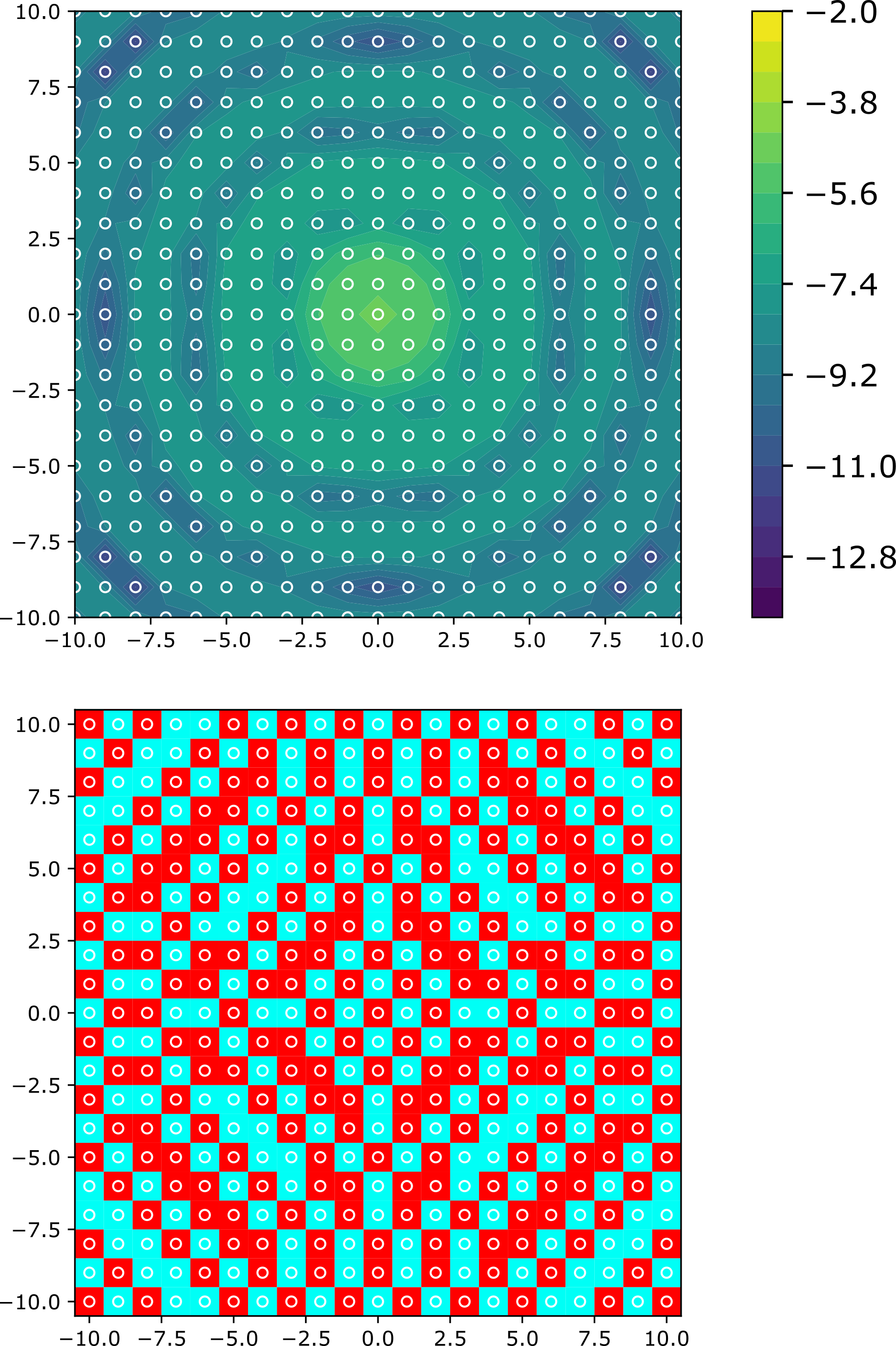}
\caption{Same as Fig. \ref{fig:x=0}, but showing the lattice results for $x=0.3, 0.45, 0.7$ and 0.9 (left to right columns). In all cases, the hopping $t=1$. }
    \label{fig:x}
\end{figure}

\begin{figure}[t]
    \centering
    \includegraphics[width=0.9\linewidth]{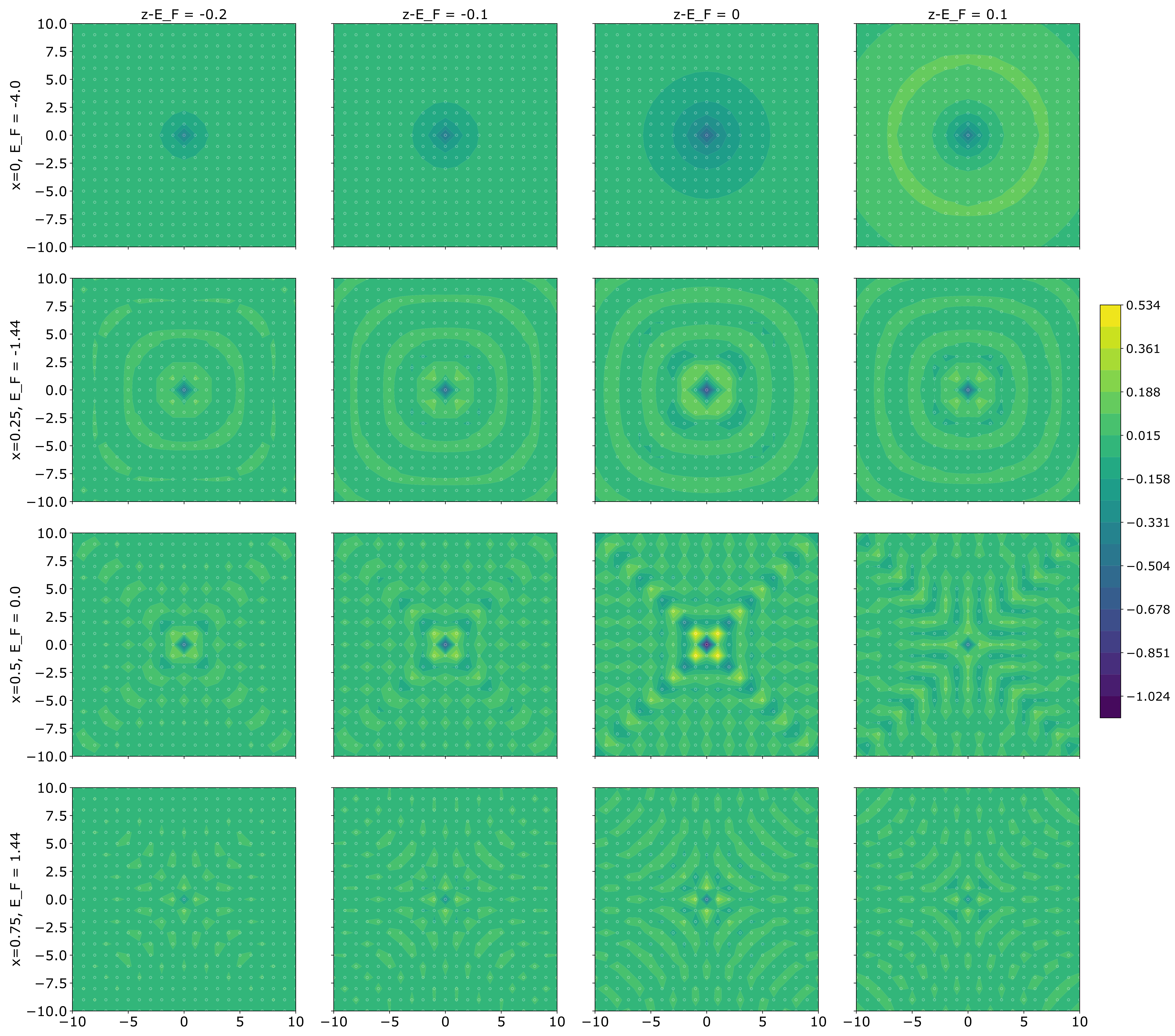}
    \caption{Maps of $\mathrm{Re} G^A_\mathbf{n}(z;x)$ versus $\mathbf{n}$. From top to bottom, the rows are for $x=0, 0.25, 0.5$ and $0.75$, respectively. The corresponding Fermi energies are $E_F=-4t, -1.44t, 0, 1.44t$. From left to right, the columns correspond to $\omega = E_F-0.2, E_F-0.1, E_F, E_F+0.1$, respectively. In all cases $t=1$ and $\eta = 0.02.$}
    \label{fig:Re}
\end{figure}

As already discussed, at energies $\omega< E_F$ the propagators are real numbers (the finite $\eta$ leads to a minute imaginary part, but this quickly becomes negligible the further $\omega$ goes  below $E_F$). As a result, we only show the magnitude and sign of $G^A_\mathbf{n}(z;x)$. These are mapped versus $\mathbf{n}$ in the two left columns of Fig. \ref{fig:x=0} for $x=0$ and $x=0.1$. The upper panels show the magnitude, on a log scale, while the bottom panels show the sign (red=negative, teal=positive). As expected, in both cases the magnitude decreases extremely fast with increasing $\mathbf{n}$, due to the lack of  eigenstates at these energies. This decrease is monotonic at $x=0$, while at $x=0.1$ there is a nearly-circular oscillation superimposed on it. This oscillation is a direct consequence of the nearly circular Fermi sea for this low-$x$ value, with the wavelength matching $\lambda_F=2\pi/k_F$. Indeed, in this low-$x$ limit the propagators agree very well with those obtained from the continuum approximation where we let the lattice constant $a \to 0$ and use a parabolic approximation for the dispersion, leading to a perfectly circular Fermi sea. The corresponding results for this continuum approximation, shown in the two right columns,  agree well with the low-$x$ lattice results.

As $x$ approaches half-filling, however, the Fermi surface distorts away from circular and the square symmetry of the underlying lattice becomes apparent in these patterns. This is confirmed in Fig. \ref{fig:x}, which displays the lattice propagators at $\omega=-10t$ for $x=0.3, 0.45, 0.7$ and 0.9. In particular, as $x\to 0.5$, the phase contour converges towards a square while the magnitude shows strong anisotropy, with significantly slower decay along the diagonal compared to that along the $x$ and $y$ axes. As $x\to 1$, on the other hand, the circular oscillation  becomes apparent again, and is now associated with the small circular hole pocket in which the electron can be injected. Its sign picks up an additional phase $\exp(i \mathbf{Q}\cdot \mathbf{R}_\mathbf{n})$, explaining the apparent 'checkerboard' pattern.

These patterns can be readily understood in the limit $|z|/t \gg 1$, where 
\begin{equation}
        G^A_{\mathbf{n}}(z;x) = \frac{1}{z}\int_{BZ} \frac{d^d\mathbf{k} }{(2\pi)^d}
    e^{i\mathbf{k}\cdot \mathbf{R}_\mathbf{n}}\Theta(\epsilon_{\mathbf{k}}-E_F) +{\cal O}\Big(\frac{t}{z^2}\Big).
    \label{14}
\end{equation}
This shows that the propagator converges towards the Fraunhofer diffraction pattern evaluated at the lattice sites' locations, with the Brillouin zone outside of the Fermi surface acting as the aperture function. Thus, for both small ($x\to 0$) and large ($x\to 1$) concentrations, the Fermi surface is approximately circular and we expect the pattern to resemble the Airy disk, with the $n^\mathrm{th}$ root of the oscillation being $R_n/k_F$, where $R_n$ is the $n^\mathrm{th}$  root of the Bessel function of order $1$. As the propagator is only meaningful at the lattice sites $\mathbf{n}$, we expect approximately that it changes sign every $\lambda_F/2$; this pattern becomes more accurate as $|\mathbf{n}|$ increases. As $x$ approaches $0.5$, the Fermi surface converges towards a square rotated by $45^\circ$ and the propagator converges towards the diffraction pattern of a square, which has a characteristic sequence of bright fringes along the diagonal and a square phase contour as the sign oscillates between positive and negative values.

At $x=0$, the integral multiplying $1/z$ in Eq. (\ref{14}) vanishes for any $\mathbf{n} \neq \mathbf{0}$, therefore $G^A_{\mathbf{n}\ne \mathbf{0}}(z;0)= {\cal O}(1/z^2)$ at large $|z|$. In contrast, at any $x\ne 0$, the integral remains finite and $G^A_{\mathbf{n}}(z;x)$ decays at a rate of $1/z$ for all $\mathbf{n}$. This implies that for $x$ not too close to $1$, $G_\mathbf{n}^A(z;x)\gg G_\mathbf{n}^A(z;0)$ for any $\mathbf{n}\ne \mathbf{0}$ and sufficiently large $|z|$, and Eq. (\ref{GeneralARrelation}) becomes $G^R_{-\mathbf{n}}(z;x) \approx -G^A_{\mathbf{n}}(z;x)$. Combining this with Eq. (\ref{bipartiteARrelation}) and that fact that $G_\mathbf{n}^A(-z;x) \approx -G_\mathbf{n}^A(z;x)$ at large $|z|$, we  find that in this limit, {\em i.e.}  for $x$ not too close to $0$ or $1$, $|z|\gg t$ and  $\mathbf{n}\ne \mathbf{0}$, $ G_\mathbf{n}^A(z; 1-x)\approx -G_\mathbf{-n}^A(z;x)e^{i\mathbf{Q}\cdot \mathbf{R}_\mathbf{n}} $, explaining the additional sign $e^{i\mathbf{Q}\cdot \mathbf{R}_\mathbf{n}}$ as $x \to 1$, in agreement with the results shown in Fig. \ref{fig:x}. 

Although the detailed analysis provided above is quantitatively valid only in the limit $|z|\gg t$, it also explains qualitatively the spatial patterns of the propagators observed for energies inside the bandwidth. As an example, Figure \ref{fig:Re} shows maps of $\mathrm{Re} G^A_\mathbf{n}(z;x)$ versus $\mathbf{n}$, for concentrations $x=0, 0.25, 0.5$ and $0.75$ (top to bottom) and energies  $\omega = E_F-0.2, E_F-0.1, E_F, E_F+0.1$ (left to right). For $z< E_F$ we see again the same patterns: circularly symmetric at $x=0$,  quasi-circular oscillations with wavelength $\lambda_F$ at $x=0.25$, square symmetry with larger values along the diagonals at $x=0.5$, and finally the additional phase $\exp(i\mathbf{Q}\cdot \mathbf{R}_\mathbf{n})$ of the $x=0.75$ pattern compared to the $x=0.25$ pattern. Once $z\ge E_F$, the patterns are dominated by the wavevectors $\mathbf{k}$ for which $\epsilon_\mathbf{k}= \omega$ [these contribute most to the integral in Eq. (\ref{addsum})] and the patterns change somewhat, although the general symmetries persist.  

\begin{figure}
    \centering
    \includegraphics[width=0.9\linewidth]{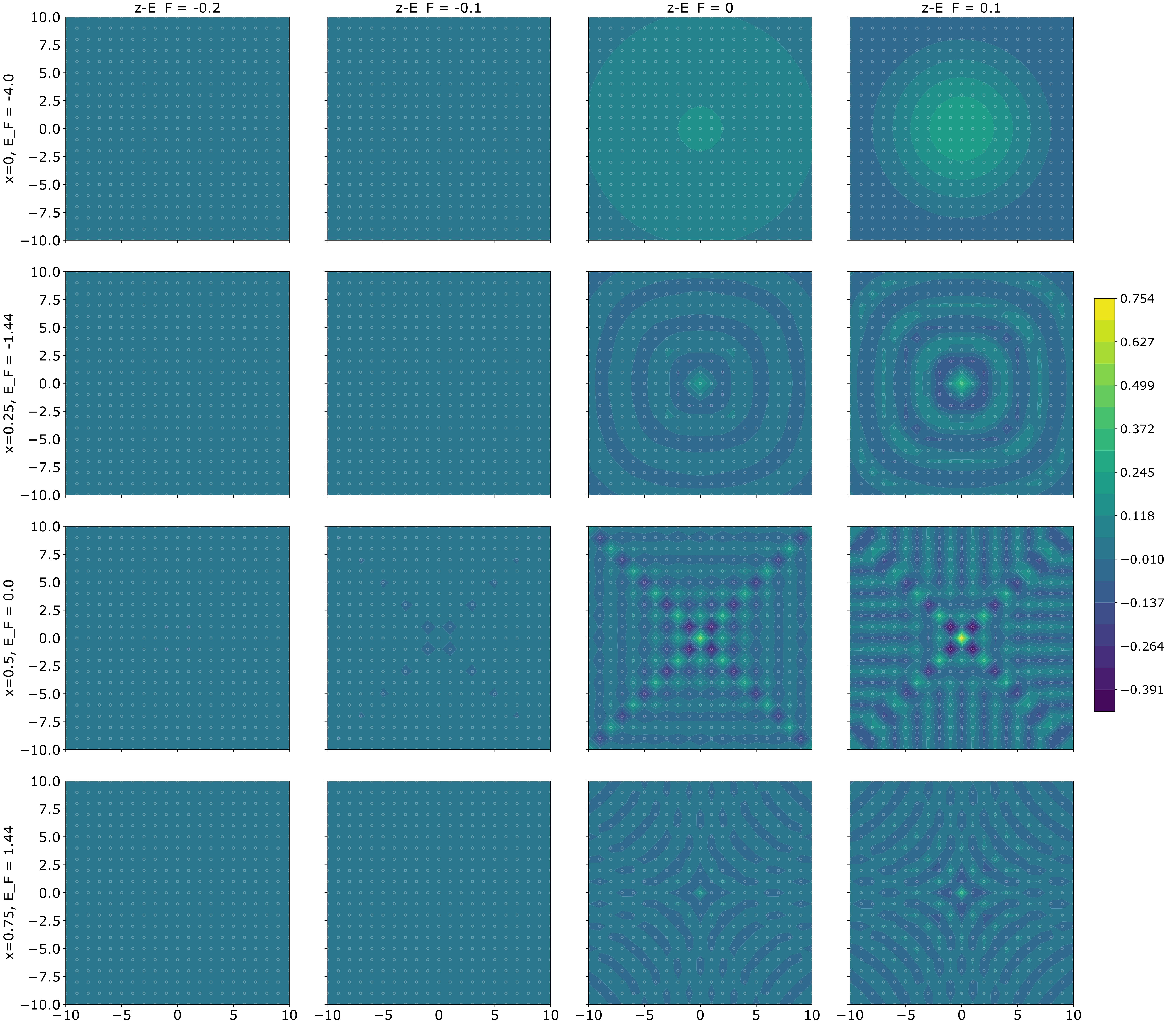}
    \caption{Same as in Fig. \ref{fig:Re} but for $-\mathrm{Im} G^A_\mathbf{n}(z;x)$. }
    \label{fig:Im}
\end{figure}

Unlike for energies outside the band, $\mathrm{Im} G^A_\mathbf{n}(z;x)$ also becomes finite for energies $E_F \le \omega < 4t$, as shown in Fig. \ref{fig:Im} for the same parameters as in Fig. \ref{fig:Re}. As discussed above,  these values are equal to $\mathrm{Im} G^A_\mathbf{n}(z;0)$; the patterns vary with $x$ simply because we measure $\omega$ from $E_F$, which changes with $x$.

\section{Discussion and Conclusions}

We have introduced an efficient scheme to compute real-space, bare lattice propagators in the presence of a Fermi sea. Because the Fermi sea imposes a sharp separation at $E_F$ between occupied and empty states, suitable combinations of the finite-$x$ addition propagators have a real or imaginary part that is trivially related to their $x=0$ counterpart, see Eqs. (\ref{intreal}) and (\ref{intimag}), while the remaining parts are recovered through Kramers-Kronig relations implemented as Hilbert transforms via fast Fourier transforms, see Eq. (\ref{13}). As a result, once the $x=0$ propagators are available, the propagators at any concentration $x$ can be generated at a cost that is independent of the lattice, the model, and the dimension $D$, in contrast to direct Brillouin-zone integration, whose cost grows rapidly with $D$. The general identity of Eq. (\ref{GeneralARrelation}) links the addition and removal propagators, halving the computational work, while Eq. (\ref{bipartiteARrelation}) provides an additional particle-hole-type relation whenever the dispersion admits a $\mathbf{Q}$ such that $\epsilon_{\mathbf{Q}-\mathbf{k}}=-\epsilon_\mathbf{k}$ for any $\mathbf{k}$.

Beyond the numerical speedup, the real-space maps have a transparent physical interpretation: at energies far outside the band, they converge to the Fraunhofer diffraction pattern whose aperture is the unoccupied part of the Brillouin zone, so the geometry of the Fermi surface -- the circular pockets at low $x$ and $1-x$, or the square contour near half-filling -- is directly imprinted on the sign structure and the decay of the propagators. This qualitative picture remains useful at energies inside the band, as illustrated in Figs. \ref{fig:Re} and \ref{fig:Im}.

These propagators are the essential input for real-space treatments of problems where translational invariance is broken by impurities, disorder, or surfaces at a finite carrier concentration, and for variational methods describing carriers dressed by bosonic clouds on top of a Fermi sea \cite{berciu2022polarons}. A natural extension is to multi-band models, where the same identities should apply to the matrix elements of the propagator, because the sharp Fermi-sea occupation factorizes for each band.  At finite temperature, on the other hand, the sharp Fermi step is replaced by a Fermi function, and the present approach no longer applies directly; whether a similarly efficient generalization exists is an interesting open question.

%\ack{Sample text inserted for demonstration.}

\funding{This project was supported by the Max Planck--UBC--UTokyo Center for Quantum Materials and by the Natural Sciences and Engineering Research Council of Canada.}

\roles{O.T. developed this method, generated the codes and the data, and wrote the original draft. M.B. provided suggestions and helped edit the final draft. }

\data{The data that support the findings of this study are available from the authors upon reasonable request.}
% For more information on IOP Publishing's research data policy see: https://publishingsupport.iopscience.iop.org/questions/research-data/

%\suppdata{Sample text inserted for demonstration.}

% \section*{References}
\bibliographystyle{apsrev4-2}
% \begin{NoHyper}
\bibliography{references}

\end{document}